# *ROSAT* Observations of the Flare Star CC Eri

H. C. Pan and C. Jordan
*Department of Physics (Theoretical Physics), University of Oxford, 1 Keble Road, Oxford OX1 3NP*

**ABSTRACT**

The flare/spotted spectroscopic binary star CC Eri was observed with the Position Sensitive Proportional Counter (PSPC) on the X-ray satellite *ROSAT* on 1990 July 9-11 and 1992 January 26-27. During the observations, the source was variable on time scales from a few minutes to several hours, with the X-ray (0.2-2 keV) luminosity in the range $\sim 2.5 - 6.8 \times 10^{29}$ erg s$^{-1}$. An X-ray flare-like event, which has a one hour characteristic rise time and a two hour decay time, was observed from CC Eri on 1990 July 10 16:14-21:34 (UT). The X-ray spectrum of the source can be described by current thermal plasma codes with two temperature components or with a continuous temperature distribution. The spectral results show that plasma at $T_e \sim 10^7$ K exists in the corona of CC Eri.

The variations in the observed source flux and spectra can be reproduced by a flare, adopting a magnetic reconnection model. Comparisons with an unheated model, late in the flare, suggest that the area and volume of the flare are substantially larger than in a solar two ribbon flare, while the electron pressure is similar. The emission measure and temperature of the non-flaring emission, interpreted as the average corona, lead to an electron pressure similar to that in a well-developed solar active region. Rotational modulation of a spot related active region requires an unphysically large X-ray flux in a concentrated area.

**Key words:** X-rays:stars – stars:individual:CC Eri – stars:late-type – stars:coronae – stars:flare – stars:rotation.

## 1 INTRODUCTION

CC Eri (HD 16157) is a spectroscopic binary with period of 1.56 days. It consists of a K7Ve (although some authors adopt M0Ve) primary and a ∼dM4 secondary with mass ratio ≈ 2 (Evans 1959, 1971; Strassmeier et al. 1993). The primary co-rotates with the orbital motion owing to the tidal lock and is one of the fastest rotating solar neighbourhood K dwarfs known. The stellar parameters adopted for the primary are given in Table 1. (See Section 4.1 for the definition of Ro, the Rossby number).

CC Eri is classed as a very active BY Draconis variable flare star (Evans 1959; Busko & Torres 1976). The system shows H Balmer and Ca II H and K lines in emission (Evans 1959). It exhibits sinusoidal optical light variations of varying amplitude and phase with a periodicity essentially identical to that of the orbital motion (Evans 1959). The variations have been attributed to the rotational modulation of dark star spots (Krzeminski 1969; Bopp & Evans 1973). The optical emission line fluxes vary roughly in anti-phase with its photometric variation (Busko, Quast & Torres 1977). The source is also visible in the far-infrared wave-band with a luminosity at 12 $\mu$m of $\sim 0.22 \times 10^{31}$ erg s$^{-1}$ (Tsikoudi 1988). Radio observations indicate that CC Eri is variable at radio wavelengths and is in a strong flaring state for certain periods of time (Caillault, Drake & Florkowski 1988; Güdel 1992).

CC Eri has also been observed at ultra-violet wavelengths with *IUE* and some evidence of flaring was noted (Byrne et al. 1992). A multi-wavelength campaign in optical and ultra-violet has shown that CC Eri is the most luminous solar neighbourhood K star yet measured in the C IV lines (Byrne et al. 1992). The observed variation of the Mg II lines with phase confirms earlier conclusions that the chromospheric emission is anti-correlated in a general sense with the optical continuum, suggesting the presence of active region emission associated with star spots.

Previous X-ray observations of CC Eri have been made with *HEAO-1*, *Einstein* and *EXOSAT* (Tsikoudi 1982; Schmitt et al. 1987, 1990; Pallavicini et al. 1988; Pallavicini, Tagliaferri & Stella 1990). These observations yield a source luminosity between 1.6 and 4.2 $\times 10^{29}$ erg s$^{-1}$ and temperatures of 4.6 to 15$\times 10^6$ K. Since none of these observations lasted very long, little information on the temporal and spectral variation of the source was available.

Taking advantage of the increased sensitivity and the low background noise of the *ROSAT* PSPC detector, compared with *Einstein* and *EXOSAT*, we have obtained high quality, low resolution spectra of CC Eri. In this paper, we present the first X-ray observations that reveal details about the spectrum of the source and its time variations. We describe the PSPC observations of CC Eri and the data analysis methods in Section 2 and present the observational results in Section 3. In Section 4 we show that the variations in the observed flux can be interpreted in terms of a flare using magnetic reconnection theory. In Section 5 we use the emission measures and temperatures to make an alternative model of the flare decay phase, assuming that *no* heating occurs, and also model the quiescent parameters. Finally in Section 6 we summarize the results, and suggest further observations.

**Table 1.** Adopted parameters of CC Eri.

| Spectral Type | $d$ (pc) | $R_\star/R_\odot$ | $\log g_\star$ (CGS) | $P$ (days) | $R_0$ | $L_\star$ (erg s$^{-1}$) | $T_{\rm eff}$ (K) |
|---|---|---|---|---|---|---|---|
| K7Ve+dM4 | 12 | 0.60 | 4.55 | 1.56 | $6 \times 10^{-2}$ | $2.80 \times 10^{32}$ | 3875 |

## 2  OBSERVATIONS AND DATA ANALYSIS

The *ROSAT* PSPC observations of CC Eri were made on 1990 July 9-11 during the *ROSAT* calibration and verification phase, and on 1992 January 26-27 on our behalf during the Guest Observer Program AO-1 phase. The Boron filter was used in the observation made on 1992 January 26-27 to allow an increase of spectral resolution at lower energies. The total exposure time was 17718 seconds, divided into 22 groups for the study of spectral variability discussed below. Table 2 lists details of the observations and the time intervals over which the X-ray spectrum was accumulated. The PSPC has an energy range of 0.1-2.4 keV with a spectral resolution of $\Delta E/E \approx 0.42$ at 1 keV, more than twice that of the *Einstein* Image Proportional Counter (IPC). A comprehensive description of the *ROSAT* satellite, the X-ray telescope and the PSPC detector can be found in Trümper (1983) and Pfeffermann et al. (1987).

The observational data were reduced using the Starlink ASTERIX X-ray data reduction package (Version 1.6b). The X-ray counts of the source were extracted from a circle, centred on the source with a radius chosen to optimize the signal-to-noise ratio (6 arcmin in the case of CC Eri). Background counts, estimated from an annulus region centred on the source cell with a larger radius (15 arcmin), but without obvious contaminating sources, were subtracted from the source counts. Standard corrections for shadowing by the wire mesh, mirror vignetting, photon scattering and also dead time were applied to the background subtracted source counts.

The X-ray pulse height spectra were analyzed using the XANADU-XSPEC package (version 8). After excluding "bad" raw pulse height analysis (PHA) channels corresponding to energies below 0.1 keV and above 2.4 keV, the 256 channel data were binned up according to the compressed channel assignment of the *ROSAT* Standard Analysis Software System (SASS). The rebinned spectrum contains 34 SASS channels and each new bin has a constant oversampling factor (2.45) of the spectral resolution. The SASS channels were selected to prevent dramatically different oversampling factors at various energies skewing the spectral fit. The latest recent version (released in 1993 January) of the *ROSAT* PSPC response matrix was used in our spectral analyses. A systematic error of one percent of the measured count rate, in each of the 34 SASS channels, was quadratically added to the photon noise in those channels to allow for uncertainties in the calibration of the PSPC response matrix. We further restrict our data analysis in the SASS channel 5-30 ($\sim 0.17 - 2$ keV) to avoid possible soft X-ray contamination (solar-scattered X-rays, auroral X-rays, etc) below 0.17 keV and the large uncertainty in the calibration (systematic error $> 10\%$) above 2 keV.

**Table 2.** CC Eri – *ROSAT* observation journal.

| Date | Start (UT) | End (UT) | Exposure (sec) | Spectral Group |
|---|---|---|---|---|
| 1990 July 9 | | | | |
| | 14:27:29 | 14:44:09 | 794 | 1 |
| | 14:44:09 | 15:00:49 | 668 | 2 |
| 1990 July 10 | | | | |
| | 16:14:09 | 16:27:29 | 663 | 3 |
| | 16:27:29 | 16:40:49 | 483 | 4 |
| | 16:47:29 | 17:07:29 | 701 | 5 |
| | 17:34:09 | 17:47:29 | 683 | 6 |
| | 17:47:29 | 18:00:49 | 800 | 7 |
| | 18:00:49 | 18:14:09 | 780 | 8 |
| | 18:20:49 | 18:34:09 | 543 | 9 |
| | 18:34:09 | 18:47:29 | 457 | 10 |
| | 19:07:29 | 19:20:49 | 433 | 11 |
| | 19:20:49 | 19:34:09 | 800 | 12 |
| | 19:34:09 | 19:47:29 | 800 | 13 |
| | 19:47:29 | 20:00:49 | 800 | 14 |
| | 20:00:49 | 20:14:09 | 800 | 15 |
| | 20:14:09 | 20:27:29 | 563 | 16 |
| | 20:47:29 | 21:04:09 | 617 | 17 |
| | 21:04:09 | 21:17:29 | 800 | 18 |
| | 21:17:29 | 21:34:09 | 931 | 19 |
| 1990 July 11 | | | | |
| | 03:40:49 | 03:51:00 | 570 | 20 |
| 1992 January 26 | | | | |
| | 21:08:31 | 21:38:31 | 1726 | 21 |
| 1992 January 27 | | | | |
| | 02:46:51 | 03:48:15 | 2306 | 22 |

Various X-ray plasma emission codes are available for modelling the X-ray spectra of late-type stars. We have adopted the most recently available versions of two of them, one by Landini and Monsignori-Fossi (LMF) and the other by Raymond and Smith (RS), to model our data in terms of the emission measures and temperatures. Discussions of these two models can be found in Landini & Monsignori-Fossi (1970, 1985, 1990) and in Raymond, Cox & Smith (1976), Raymond & Smith (1977) and Raymond (1988). Both models assume an optically thin plasma in statistical equilibrium and include emission from important elements in addition to continuum radiation from bremsstrahlung, recombination and two-photon processes. Given the low spectral resolution of our PSPC data we have frozen elemental abundances at values given by Anders & Grevesse (1989) and Grevesse, Noels & Sauval (1992) for the solar photosphere. We have not attempted to model the spectral lines in detail.

When applying the LMF and RS codes to fit our data, we assume that the X-ray emission from the star can be described by the form

$$F(T_e, E) = \frac{1}{4\pi d^2} \int N_e^2 dV \, P(T_e, E) \text{ erg cm}^{-2} \text{ s}^{-1} \text{ keV}^{-1} \quad (1)$$

where $d$ is the source distance, $N_e$ is the electron density, and $P(T_e, E)$ is the plasma emissivity as given by the LMF or RS codes at electron temperature $T_e$ and energy $E$. The volume emission measure, $Em(V) = \int N_e^2 dV$, derived, assumes that all photons escape, none being intercepted by the presence of the star. This is referred to later as the "apparent" emission measure.

We have used several different combinations of the RS and LMF models to fit the PSPC spectra to derive temperatures and emission measures by minimizing the $\chi^2$ statistic. These combinations include one, two, and multiple temperature models (i.e. the X-ray emission is $\sum_{i=1}^{n} F(T_i, E)$, $n = 1, 2, 3...$), and a continuous emission measure distribution (based on the RS plasma code) in which the volume emission measure at a given temperature $T_e$ is defined as $Em(V, T_e) = Em(V)_{\max}(T_e/T_{\max})^\alpha$, where $Em(V)_{\max}$ is the volume emission measure at the maximum temperature $T_{\max}$, and $\alpha$ is the gradient. Using the differential emission measure $dEm(V, T_e)/dT_e = N_e^2 dV/dT_e$, the X-ray flux from the star in the continuous emission measure distribution model can be expressed as

$$F(T_{\max}, E) = \frac{\alpha \, Em(V)_{\max}}{4\pi d^2} \frac{1}{T_{\max}^\alpha}$$
$$\int_{T_{\min}}^{T_{\max}} T_e^{\alpha-1} P(T_e, E) dT_e \text{ erg cm}^{-2} \text{ s}^{-1} \text{ keV}^{-1} \quad (2)$$

where $T_{\min}$ is the minimum temperature. In the Sun the emission measure distribution as a function of $T_e$ has a roughly constant gradient down to $T_e \sim 2 \times 10^5$ K, so this has been adopted for $T_{\min}$. However, the spectral range included by the PSPC contains little emission from lines formed below $4.7 \times 10^5$ K, and the consequences of using this higher $T_{\min}$ are discussed in Section 3.2.

It is important to stress that the one, two, and multiple temperature models provide only a parameterization of the coronal emission and do not necessarily represent a physically consistent description of the physical processes in the emitting plasma. Observations of the quiet Sun and solar active regions show that a continuous emission measure distribution is present (e.g. Malinovsky & Heroux 1973; Dere 1982).

We find that the one-temperature model gives an unacceptable fit to all of the PSPC spectra and that including more than two temperatures does not improve the statistics of the spectral fit. Therefore these two types of models will not be discussed any further in this paper except when presenting the *Einstein* IPC results where the 1T model gives an adequate description to the low spectral resolution IPC data. In general the two-temperature (2T) RS and LMF model and the continuous emission measure (CEM) RS model give equally good fits to the PSPC spectra.

Interstellar absorption is considered using the model of Morrison & McCammon (1983). However, the hydrogen column density $N_H$, when allowed to vary as a free parameter in the fit for the 2T and CEM models, is not well constrained by the compressed 0.17-2 keV PSPC spectrum. We have therefore fitted the 2T and CEM model to the data with column density $N_H$ fixed at 0.0, 2.6, 10, and 29 $\times 10^{18}$ cm$^{-2}$. All the models give an acceptable fit to the data. We note that varying $N_H$ in the range of 0.0-29$\times 10^{18}$ cm$^{-2}$ has little or no effect on the temperature and emission measure of the high temperature component of the 2T model. However it causes an average change of $\sim 10\%$ in the temperature (which decreases as $N_H$ increases) and in the emission measure (which increases as $N_H$ increases) of the low temperature component. These small changes, in most cases, are within the one-sigma uncertainties of the temperature and emission measure (see Table 3). For the CEM model, both the gradient $\alpha$ and the emission measure are correlated with the change of the low energy absorption and the temperature $T_{\max}$ does not change with $N_H$. Since $N_H = 2.6 \times 10^{18}$ cm$^{-2}$ is estimated from an average interstellar hydrogen density of 0.07 cm$^{-3}$ (Paresce 1984) and the distance of CC Eri, in the following sections we will only present the spectral results derived with this column density.

## 3 RESULTS

### 3.1 X-ray Intensity Variability

CC Eri was variable on time scales from a few minutes to several hours during the PSPC observations. The X-ray (0.2-2 keV) luminosity of the source was in the range $\sim 2.5 - 6.8 \times 10^{29}$ erg s$^{-1}$. We have re-analyzed the X-ray observations of CC Eri made with the *Einstein* IPC and the *EXOSAT* Low Energy (LE) telescope in order to compare our *ROSAT* results with them in a similar energy range. The IPC measurement gives a luminosity of $\sim 2.6 \times 10^{29}$ erg s$^{-1}$ in the 0.2-2 keV band and the *EXOSAT* LE gives $\sim 6.5 \times 10^{29}$ erg s$^{-1}$ in the same energy band, both of which lie within the range of values from the *ROSAT* observations.

Fig. 1 shows the X-ray light curve of CC Eri obtained on 1990 July 10 16:14-21:34 (UT). The data are in segments separated either by earth occultation of the source or portions of orbits with high background. A flare-like event was observed in this period. The source became gradually brighter and reached a peak flux of $6.3 \pm 0.1$ counts s$^{-1}$ (corresponding to a luminosity $\sim 6.5 \times 10^{29}$ erg s$^{-1}$) at around 17:53 (UT). The source flux slowly decreased again and after about 3.7 hours the luminosity became $\sim 3.3 \times 10^{29}$ erg s$^{-1}$. There is a bump in the light curve between 19:39-19:57. If the source behaviour did not change dramatically during the gaps, the observed flare-like event has a one hour (or less) characteristic rise time and a two hour decay time. The ratio of the peak to minimum fluxes is about 2.

### 3.2 X-ray Spectrum Variability

Figs. 2(a) and 2(b) show respectively the two X-ray spectra of CC Eri, obtained on 1990 July 10 17:47:29-18:00:49 (spectral group 7) when the source was at the peak luminosity, and on 1992 January 26 21:08:31-21:38:31 (spectral group 21). The boron filter was used in obtaining the spectrum shown in Fig. 2(b). The spectra are shown in terms of the detector counts (crosses). The fits with the 2T RS model are shown as histograms. The residuals of the model are also given in Fig. 2 in terms of the significance which is defined as (data−model)/one-sigma error. It is obvious from Fig. 2 that the 2T RS model gives a good fit to the spectra. The fits using the CEM RS code are not shown, but are virtually indistinguishable in the fits and residuals. The parameters of all 22 PSPC spectra are listed in Table 3 for the 2T RS and LMF models and in Table 4 for the CEM RS model. $T_1$, $Em(V)_1$, $T_2$ and $Em(V)_2$ in Table 3 are the temperatures and volume emission measures of the first and second components. The spectral results from fitting the *Einstein* IPC spectrum with the 1T RS & LMF models and the CEM RS model are also listed in Table 3-4 for the purpose of comparison. The errors quoted in both the tables and figures discussed below correspond to a 68% confidence level.

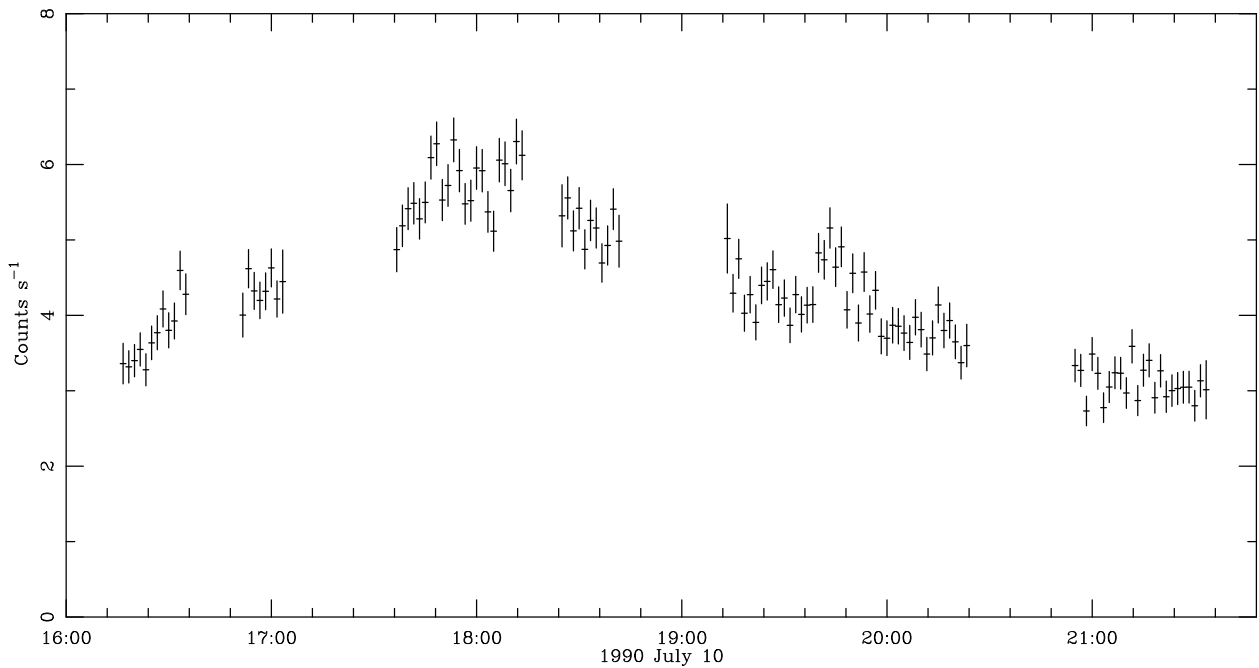

**Figure 1.** X-ray light curve of CC Eri on 1990 July 10. Each data point represents the (0.2-2 keV) X-ray flux (counts s$^{-1}$) in a 100 second time bin. The one sigma error is given for each data point.

Since the available codes give different results we have examined the relation between the model parameters produced. As can be seen from the numbers given in Table 3 there are systematic differences between the spectral results derived with the RS and LMF plasma codes although they both give adequate fits to the spectra and give a similar relationship between the source spectral parameters and its flux. The emission measures of the high temperature component are correlated with the source flux. Because the LMF code leads to larger high temperature emission measures, the smaller, low temperature component is less systematic in its behaviour. The temperatures derived from both the RS and LMF models cover a smaller range of values than do the emission measures. The variation of $T_2$ from the RS code (but not from the LMF code) increases slightly with the X-ray flux. The temperature $T_1$ does not vary systematically with the flux and has an average level of $\sim 2.3 \times 10^6$ K for the RS model and with $\sim 1.8 \times 10^6$ K for the LMF model. The IPC temperatures and emission measures given in Table 3 are within the range of the PSPC measurements.

There is a clear correlation between the ratio of the high temperature emission measures and the temperature $T_2$, as shown in Figure 3. $Em(V)_2(\text{LMF})/Em(V)_2(\text{RS})$ decreases systematically as either $T_2(\text{LMF})$ or $T_2(\text{RS})$ increases. The situation regarding the low temperature component, however, is more complex. In most cases for the same PSPC spectrum when the LMF model gives a lower temperature it will at the same time yield a lower emission measure than the RS model. These systematic differences demonstrate the systematic, temperature dependent uncertainties in the plasma emissivity codes currently available, e.g. in the ion populations adopted.

The numbers given in Table 4 show that the plasma temperature $T_{\max}$ of the CEM RS model is in the range of $11-32 \times 10^6$ K and is roughly correlated with the source flux. The emission measure $Em(V)_{\max}$ shows a small increase with the flux (Note that the two highest values of $Em(V)_{\max}$ come from spectra obtained using the Boron filter). The gradient $\alpha$, which does not correlate with the flux, is spread around a weighted average value of $1.15 \pm 0.02$. We have fitted the *Einstein* IPC spectrum using the CEM RS model with $N_H$ fixed at $2.6 \times 10^{18}$ cm$^{-2}$ and $\alpha$ at 1.15. As given in Table 4 these measurements of $T_{\max}$ and $Em(V)_{\max}$ are consistent with the PSPC results. The spectral results from the CEM RS model depend on the value of the minimum temperature $T_{\min}$. Changing $T_{\min}$ from $2 \times 10^5$ K to $4.7 \times 10^5$ K does not change $T_{\max}$ and $N_H$, but would increase $Em(V)_{\max}$ by $\sim 40\%$ and $\alpha$ by $\sim 10\%$. Given the present uncertainties in the codes we do not at this stage attempt to interpret the value of $\alpha$.

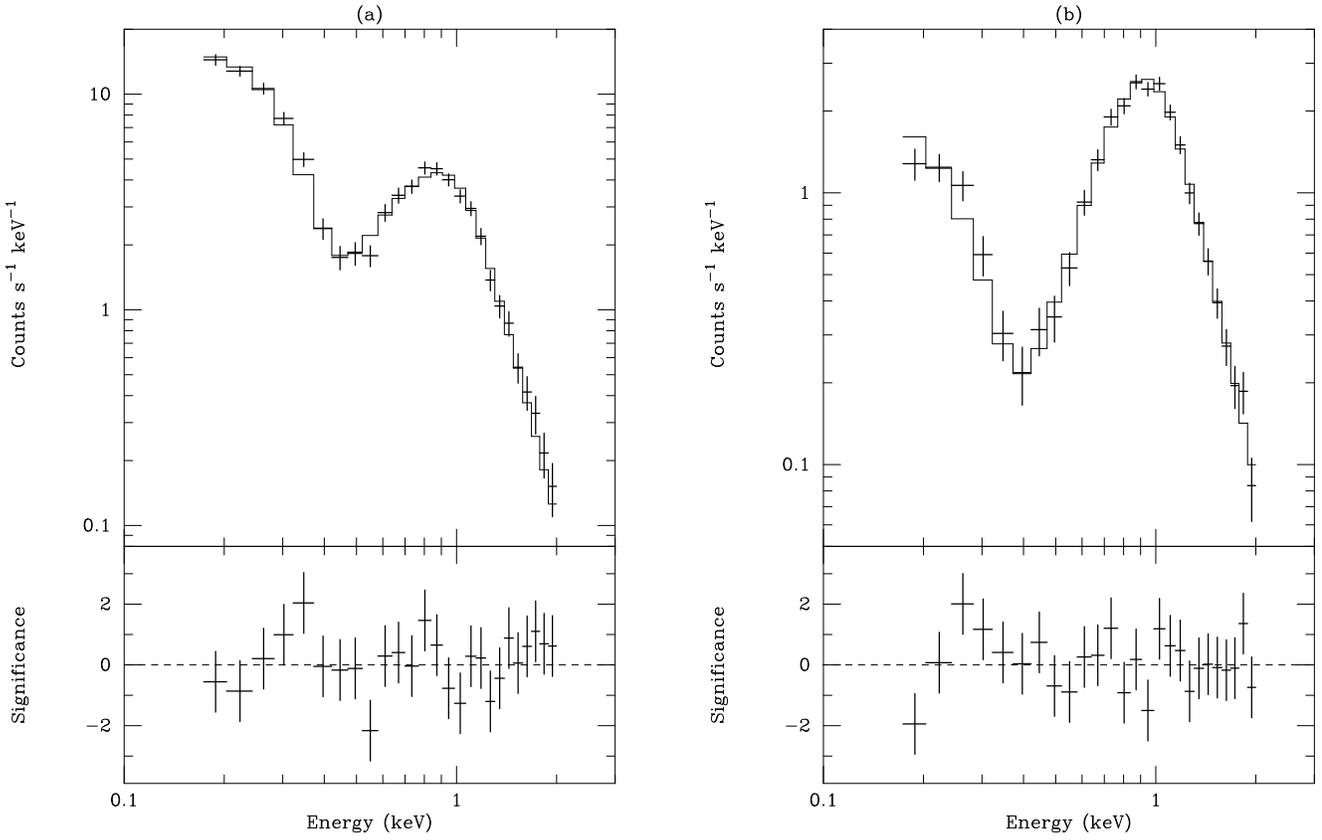

**Figure 2.** X-ray spectra of CC Eri fitted with the 2T RS model. (a) Data obtained on 1990 July 10 17:47:29–18:00:49; (b) Observation on 1992 January 26 made with the Boron filter. The top panels show the spectra in terms of the detector counts (crosses, in units of counts s$^{-1}$ keV$^{-1}$), and the 2T model (solid histogram). The residuals of the model fitting are given at the bottom of (a) & (b).

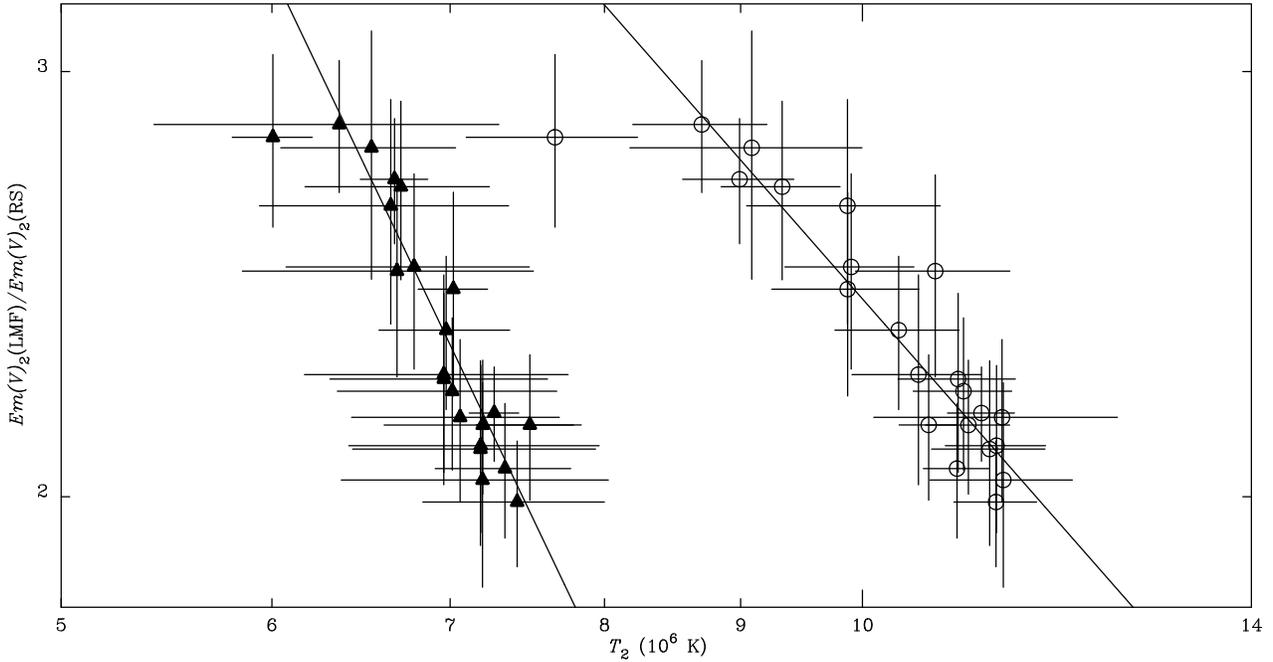

**Figure 3.** Correlation between $Em(V)_2$(LMF)/$Em(V)_2$(RS) and $T_2$(LMF) (filled triangles) or $T_2$(RS) (open circles). The solid lines represent the least square fits to the data points.

**Table 3.** Spectral parameters of CC Eri (2T models[1]).

| Spectral Group | $T_1$ ($10^6$ K) | $Em(V)_1$ ($10^{52}$ cm$^{-3}$) | $T_2$ ($10^6$ K) | $Em(V)_2$ ($10^{52}$ cm$^{-3}$) | $F_x^{[2]}$ | $\chi_r^2(dof)$ |
|---|---|---|---|---|---|---|
| 1 | 2.25 ± 0.24 | 0.68 ± 0.10 | 8.70 ± 0.50 | 1.10 ± 0.05 | 2.40 | 1.2(22) |
|   | 1.80 ± 0.97 | 0.21 ± 0.12 | 6.36 ± 0.94 | 3.13 ± 0.12 | 2.28 | 1.3(22) |
| 2 | 2.52 ± 0.29 | 0.61 ± 0.09 | 9.87 ± 0.82 | 0.98 ± 0.09 | 1.96 | 0.7(22) |
|   | 2.04 ± 0.58 | 0.18 ± 0.11 | 6.65 ± 0.71 | 2.58 ± 0.14 | 1.85 | 0.8(22) |
| 3 | 2.35 ± 0.21 | 0.67 ± 0.08 | 10.50 ± 0.59 | 1.21 ± 0.10 | 2.21 | 0.5(22) |
|   | 2.15 ± 0.32 | 0.40 ± 0.12 | 6.96 ± 0.79 | 2.73 ± 0.16 | 2.06 | 0.7(22) |
| 4 | 2.13 ± 0.19 | 0.59 ± 0.07 | 9.87 ± 0.63 | 1.46 ± 0.13 | 2.47 | 1.9(22) |
|   | 1.24 ± 0.47 | 0.13 ± 0.09 | 7.02 ± 0.21 | 3.55 ± 0.13 | 2.36 | 1.8(22) |
| 5 | 2.76 ± 0.42 | 0.81 ± 0.13 | 11.28 ± 1.18 | 1.76 ± 0.11 | 2.76 | 1.0(22) |
|   | 1.97 ± 0.55 | 0.19 ± 0.11 | 7.06 ± 0.63 | 3.80 ± 0.16 | 2.58 | 1.0(22) |
| 6 | 2.19 ± 0.14 | 0.76 ± 0.07 | 10.85 ± 0.31 | 2.36 ± 0.14 | 3.39 | 0.9(22) |
|   | 1.60 ± 0.33 | 0.25 ± 0.08 | 7.34 ± 0.43 | 4.85 ± 0.14 | 3.23 | 0.8(22) |
| 7 | 2.47 ± 0.25 | 0.94 ± 0.11 | 11.23 ± 0.40 | 2.61 ± 0.13 | 3.76 | 1.0(22) |
|   | 2.04 ± 0.38 | 0.36 ± 0.13 | 7.42 ± 0.58 | 5.19 ± 0.18 | 3.54 | 0.9(22) |
| 8 | 2.55 ± 0.23 | 0.96 ± 0.11 | 10.96 ± 0.40 | 2.40 ± 0.13 | 3.70 | 2.1(22) |
|   | 2.13 ± 0.46 | 0.27 ± 0.13 | 7.20 ± 0.59 | 5.15 ± 0.19 | 3.49 | 2.2(22) |
| 9 | 2.70 ± 0.28 | 0.93 ± 0.12 | 11.23 ± 0.49 | 2.16 ± 0.14 | 3.32 | 1.6(22) |
|   | 2.21 ± 0.48 | 0.30 ± 0.15 | 7.19 ± 0.78 | 4.53 ± 0.21 | 3.12 | 1.7(22) |
| 10 | 2.53 ± 0.29 | 0.88 ± 0.13 | 11.16 ± 0.55 | 2.05 ± 0.15 | 3.17 | 1.6(22) |
|    | 1.95 ± 0.43 | 0.32 ± 0.15 | 7.19 ± 0.75 | 4.28 ± 0.21 | 2.96 | 1.6(22) |
| 11 | 2.53 ± 0.30 | 0.79 ± 0.12 | 11.30 ± 0.70 | 1.79 ± 0.14 | 2.76 | 1.2(22) |
|    | 1.96 ± 0.41 | 0.35 ± 0.15 | 7.20 ± 0.83 | 3.64 ± 0.21 | 2.56 | 1.3(22) |
| 12 | 2.28 ± 0.24 | 0.70 ± 0.09 | 10.32 ± 0.55 | 1.56 ± 0.10 | 2.65 | 1.1(22) |
|    | 1.60 ± 0.30 | 0.21 ± 0.07 | 6.98 ± 0.39 | 3.65 ± 0.11 | 2.49 | 0.9(22) |
| 13 | 2.60 ± 0.24 | 0.86 ± 0.10 | 10.91 ± 0.46 | 1.76 ± 0.10 | 2.93 | 0.9(22) |
|    | 2.13 ± 0.34 | 0.31 ± 0.12 | 7.01 ± 0.66 | 3.88 ± 0.16 | 2.74 | 1.3(22) |
| 14 | 2.49 ± 0.23 | 0.83 ± 0.09 | 10.86 ± 0.55 | 1.46 ± 0.10 | 2.59 | 1.2(22) |
|    | 2.04 ± 0.33 | 0.42 ± 0.13 | 6.96 ± 0.65 | 3.26 ± 0.17 | 2.41 | 1.4(22) |
| 15 | 2.10 ± 0.19 | 0.63 ± 0.08 | 8.99 ± 0.43 | 1.15 ± 0.05 | 2.32 | 1.4(22) |
|    | 1.65 ± 0.45 | 0.25 ± 0.10 | 6.67 ± 0.19 | 3.10 ± 0.11 | 2.21 | 1.9(22) |
| 16 | 2.26 ± 0.19 | 0.66 ± 0.08 | 9.33 ± 0.48 | 1.16 ± 0.08 | 2.34 | 1.1(22) |
|    | 1.75 ± 0.49 | 0.21 ± 0.11 | 6.71 ± 0.53 | 3.13 ± 0.16 | 2.20 | 1.5(22) |
| 17 | 1.91 ± 0.19 | 0.48 ± 0.07 | 7.66 ± 0.57 | 0.81 ± 0.05 | 1.84 | 1.6(22) |
|    | 1.46 ± 0.26 | 0.22 ± 0.06 | 6.01 ± 0.21 | 2.28 ± 0.11 | 1.76 | 1.9(22) |
| 18 | 2.80 ± 0.29 | 0.69 ± 0.09 | 10.65 ± 0.71 | 1.05 ± 0.08 | 2.04 | 1.0(22) |
|    | 2.30 ± 0.38 | 0.23 ± 0.11 | 6.69 ± 0.84 | 2.61 ± 0.14 | 1.91 | 1.2(22) |
| 19 | 2.33 ± 0.18 | 0.57 ± 0.06 | 9.90 ± 0.55 | 0.97 ± 0.07 | 1.90 | 0.9(22) |
|    | 2.05 ± 0.35 | 0.26 ± 0.10 | 6.79 ± 0.71 | 2.41 ± 0.13 | 1.77 | 1.5(22) |
| 20 | 2.18 ± 0.28 | 0.51 ± 0.09 | 9.09 ± 0.91 | 0.68 ± 0.05 | 1.54 | 0.6(22) |
|    | 1.72 ± 0.59 | 0.27 ± 0.19 | 6.54 ± 0.49 | 1.89 ± 0.17 | 1.44 | 0.6(22) |
| 21 | 2.09 ± 0.16 | 1.09 ± 0.10 | 10.59 ± 0.27 | 2.39 ± 0.09 | 3.93 | 1.0(22) |
|    | 1.55 ± 0.45 | 1.22 ± 0.29 | 7.50 ± 0.34 | 5.13 ± 0.30 | 3.94 | 1.1(22) |
| 22 | 2.43 ± 0.20 | 1.21 ± 0.12 | 11.08 ± 0.32 | 2.22 ± 0.08 | 3.80 | 1.9(22) |
|    | 1.51 ± 0.33 | 1.16 ± 0.25 | 7.27 ± 0.16 | 4.80 ± 0.13 | 3.72 | 1.6(22) |
| IPC[3] |  |  | 10.32 ± 0.86 | 1.07 ± 0.15 | 1.51 | 0.5(8) |
|        |  |  | 8.99 ± 0.86 | 2.43 ± 0.11 | 1.53 | 0.3(8) |

[1] Upper entries – 2T RS model; and lower entries – 2T LMF model.
[2] 0.2-2 keV flux in unit of $10^{-11}$ erg cm$^{-2}$ s$^{-1}$.
[3] 1T RS & LMF model fits to *Einstein* IPC data.

Table 4. Spectral parameters of CC Eri (CEM RS-model).

| Spectral Group | $T_{\max}$ ($10^6$ K) | $\alpha$ | $Em(V)_{\max}$ ($10^{52}$ cm$^{-3}$) | $F_x^{\mathbf{1}}$ | $\chi_r^2(dof)$ |
|---|---|---|---|---|---|
| 1 | $13.4 \pm 1.1$ | $1.03 \pm 0.06$ | $2.4 \pm 0.5$ | 2.24 | 1.3(23) |
| 2 | $16.1 \pm 1.0$ | $1.09 \pm 0.13$ | $2.4 \pm 1.3$ | 1.83 | 0.7(23) |
| 3 | $20.2 \pm 1.3$ | $1.22 \pm 0.07$ | $4.5 \pm 1.0$ | 2.06 | 0.7(23) |
| 4 | $19.1 \pm 2.5$ | $1.04 \pm 0.08$ | $3.1 \pm 1.0$ | 2.40 | 1.7(23) |
| 5 | $22.4 \pm 2.4$ | $1.00 \pm 0.14$ | $3.2 \pm 1.6$ | 2.61 | 0.9(23) |
| 6 | $27.6 \pm 2.6$ | $1.12 \pm 0.14$ | $6.9 \pm 4.2$ | 3.31 | 0.7(23) |
| 7 | $31.6 \pm 3.5$ | $1.11 \pm 0.10$ | $7.3 \pm 4.1$ | 3.63 | 0.7(23) |
| 8 | $24.8 \pm 2.0$ | $0.98 \pm 0.06$ | $4.3 \pm 1.0$ | 3.55 | 1.9(23) |
| 9 | $22.6 \pm 2.4$ | $1.00 \pm 0.12$ | $3.8 \pm 1.9$ | 3.14 | 1.7(23) |
| 10 | $26.5 \pm 3.8$ | $1.12 \pm 0.09$ | $5.8 \pm 1.7$ | 3.01 | 1.5(23) |
| 11 | $25.2 \pm 1.8$ | $1.17 \pm 0.08$ | $5.8 \pm 1.6$ | 2.60 | 1.3(23) |
| 12 | $21.3 \pm 1.0$ | $1.11 \pm 0.09$ | $4.1 \pm 1.3$ | 2.53 | 0.9(23) |
| 13 | $21.3 \pm 1.1$ | $1.10 \pm 0.06$ | $4.3 \pm 1.6$ | 2.74 | 1.3(23) |
| 14 | $26.2 \pm 0.8$ | $1.30 \pm 0.07$ | $7.8 \pm 3.4$ | 2.43 | 1.3(23) |
| 15 | $14.8 \pm 1.5$ | $1.10 \pm 0.12$ | $3.0 \pm 0.7$ | 2.21 | 1.6(23) |
| 16 | $15.3 \pm 1.4$ | $1.07 \pm 0.09$ | $2.7 \pm 0.7$ | 2.19 | 1.3(23) |
| 17 | $11.2 \pm 1.0$ | $1.08 \pm 0.16$ | $2.0 \pm 1.3$ | 1.76 | 1.6(23) |
| 18 | $21.3 \pm 6.7$ | $1.21 \pm 0.03$ | $4.0 \pm 1.8$ | 1.89 | 1.3(23) |
| 19 | $21.2 \pm 0.8$ | $1.25 \pm 0.33$ | $4.2 \pm 1.6$ | 1.78 | 1.4(23) |
| 20 | $14.8 \pm 1.3$ | $1.28 \pm 0.09$ | $3.0 \pm 0.8$ | 1.45 | 0.6(23) |
| 21 | $22.0 \pm 0.7$ | $1.61 \pm 0.15$ | $41.4 \pm 19.4$ | 3.59 | 1.4(23) |
| 22 | $23.4 \pm 1.1$ | $1.36 \pm 0.11$ | $15.3 \pm 7.0$ | 3.52 | 1.6(23) |
| IPC | $22.0 \pm 1.4$ | 1.15 | $3.1 \pm 0.4$ | 1.45 | 0.9(8) |

$^{\mathbf{1}}$ 0.2-2 keV flux in unit of $10^{-11}$ erg cm$^{-2}$ s$^{-1}$.

Even for the bins excluding the flare, the spectral results derived with either the 2T or CEM RS model all show the presence of high temperature ($\sim 10^7$ K) plasma around CC Eri. This is consistent with the previous detections of a number of flare stars in quiescence with the *EXOSAT* ME and the *Einstein* IPC (e.g. Pallavicini et al. 1990; Schmitt et al. 1990).

## 4  DISCUSSION

### 4.1  The 1990 July 10 Event – the Counterpart of a Solar Two Ribbon Flare?

Pallavicini et al. (1990) classify the 32 flares observed on dMe stars with the *EXOSAT* LE detector as *impulsive flares* and *long decay flares* according to their X-ray decay times. Those in the first group, which are similar to solar compact flares, have rapid rise and decay times. Those in the second group, which are similar to solar two ribbon flares, have decay times of roughly one hour or longer. However the X-ray luminosity and the total energy released during the flares of both types are larger than their solar analogs by several orders of magnitude. The very energetic stellar flares have volumes and inferred magnetic field strengths which are much larger than their solar counterparts.

The *ROSAT* observations presented in the previous section indicate that the X-ray spectrum of CC Eri is variable and that the temperature and emission measure of the corona plasma to some extent rise and decay with the source intensity. Such a correlation could imply that a flare is present and the corona plasma is being heated during the flare rise and cooling during the decay phase. The event detected on 1990 July 10 from CC Eri (see Fig. 1), with a one hour rise time and a two hour decay time, may be a counterpart of a solar two ribbon flare. The energy released during the flare in the X-ray passband alone is $\sim 2.9 \times 10^{33}$ erg, after subtracting the post-flare quiescent flux, which is an order of magnitude larger than the *total* energy in a typical large solar flare ($\sim 10^{32}$ erg over $\sim 10^4$ s, e.g. Priest 1981).

In solar two ribbon flares an entire arcade of magnetic loops is disrupted by an eruptive event created by the complex thermal and non-thermal processes which occur during the flare rise phase. This open field configuration subsequently closes back, leading to the formation of a growing system of magnetic loops whose footpoints are anchored in the bright H$_\alpha$ ribbons. It is believed that the growing system of loops is a consequence of magnetic reconnection (Forbes, Malherbe & Priest 1989). The reconnection theory assumes that the open field structure relaxes back to a closed configuration of lower energy and the excess magnetic energy released in this merging process appears as the thermal energy of the bright X-ray loops (see Forbes et al. 1989 for recent development in the theory). For example, based on this hypothesis an analytical model for the reconnection process was developed by Kopp & Poletto (1984) and was applied with some success to the decay phase of several solar two ribbon flares. Such a time dependent model has also been applied to long duration stellar flares observed from the flare stars EQ Peg and Prox Cen with the *Einstein* and *EXOSAT* observatories (Poletto, Pallavicini & Kopp 1988). It was shown that the model is capable of reproducing the energy release rate and temporal evolution during the flare decay phase. Given a number of assumptions a set of well-defined physical parameters for the emitting regions of EQ Peg and Prox Cen was derived by fitting the analytical model to the data.

In order to investigate the physical nature of the flare on CC Eri, we have adopted the magnetic reconnection theory of Kopp & Poletto (1984) to model the decay phase of the event on 1990 July 10. Other flare models could be chosen, and here we show only that this particular model leads to solutions with physically plausible parameters. We have made the same sets of assumptions as in Poletto et al. (1988) regarding the location and size of the active region. Details of the model and the assumptions made can be found in Kopp & Poletto (1984) and in Poletto et al. (1988). Briefly, the energy release rate per radian of longitude during the magnetic reconnection is given by:

$$\frac{dE}{dt} = (1/8\pi) 2n(n+1)(2n+1)^2 R_\star^3 B_m^2$$
$$[I_{1,2}(n)/P_n^2(\theta_{1,2})] \frac{y^{2n}[y^{(2n+1)} - 1]}{[n + (n+1)y^{(2n+1)}]^3}(dy/dt) \quad (3)$$

where $I_{1,2} = \int P_n^2(\theta) d(\cos\theta)$, and $P_n(\theta)$ is the Legendre polynomial of degree n, which is 5, 9, 17 and 35, corresponding to the latitude width of the region 33°, 20°, 10°, and 5° respectively. The longitude width is assumed to be 1.5 times the latitude width, as typically seen in solar two ribbon flares. $B_m$ is the maximum surface field within the region and $R_\star$ is the stellar radius of CC Eri, which is estimated to be $\sim 0.60 R_\odot$ from the absolute visual magnitude given by Evans (1959) and Pettersen (1991). The time dependent function y is of the form $y(t) = 1 + (H_m/R_\star)[1 - \exp(-t/t_0)]$. The parameter $H_m$ is the maximum height reached by the reconnection point during its upward movement and is assumed to be equal to the separation of the loop footpoints. $t_0$ is the time constant of the energy release, which is found from the modelling and has a mean value of 17 ks. Following Poletto et al. (1988) it is assumed that 10 percent of the energy released by the magnetic reconnection process is radiated in the X-ray band.

To estimate the energy release rate during the 1990 July 10 event, we have fitted the CEM RS model to the difference spectra of CC Eri, with $N_H$ fixed at $2.6 \times 10^{18}$ cm$^{-2}$ and $\alpha$ at 1.15. The X-ray light curve shown in Fig. 1 has been subdivided into 17 time intervals (labelled spectral group 3-19 in Table 2) over which the PHA spectra have been accumulated. The difference spectra are obtained by subtracting the quiescent spectrum measured in group 19 from the spectra 3-18. As given in Table 5, in the decay phase of the 1990 July 10 event, the temperature decreases from $2.8 \times 10^7$ K to $1.2 \times 10^7$ K and the emission measure from $4.2 \times 10^{52}$ to $0.1 \times 10^{52}$ cm$^{-3}$.

We have fitted equation (3), after integration along the longitude and multiplied by a constant 0.1, to the energy release rate measured with the PSPC during the decay phase. In Fig. 4 we plot the observed energy release rate together with the best fit curve which has little dependence on the polynomial degree n. It is obvious that this magnetic reconnection model could account for the decay phase of the flare-like event on 1990 July 10 except in the period 19:39-19:57 where a second bump exists. (This may be caused by a second flare). In Table 6, we list a set of geometrical and physical parameters of the X-ray emitting region derived from the model fitting. The maximum surface magnetic field within the region $B_m$ ranges from 252 Gauss to 1474 Gauss, depending on the size of the flare region. This is the *minimum* value required to account for the flare energy. $V_{rise}$, the initial upward velocity of the reconnection point, defined as $V_{rise} = H_m/t_0$, is similar to those derived for EQ Peg and Prox Cen (Poletto et al. 1988). The electron density $N_e$, which is evaluated for the decay phase from the observed emission measure (see Table 5) by approximating the X-ray emitting loop arcade with a semi-cylinder whose volume at time $t$ is $\sim \frac{1.5\pi H_m^3}{2}[1 - \exp(-t/t_0)]^2$, decreases by about an order of magnitude from the flare peak to the quiescent state. The values of $N_e$ and $B_m$ required lie between those found for the flares on EQ Peg and Prox Cen (Polleto et al. 1988), but the values of $H_m$ are larger because the radius of CC Eri is larger. The last column in Table 6 represents a limiting case since it gives a radiation loss (0.2-2 keV band) that just exceeds the stellar surface flux.

There are no measurements of the surface magnetic field of CC Eri. Measurements for other stars suggest that the fields are close to the equipartition values (Saar, Linsky & Beckers 1986). Using $T_{eff}$ as given in Table 1, the equipartition field would be 2630 Gauss. The observations of 17 stars with measured fields and area filling factors (see Montesinos & Jordan 1993) give a relation between log $f_s$ and the Rossby number $Ro$ ($Ro = P_{rot}/\tau_c$, where $\tau_c$ is the turnover time at the base of the convection zone, calculated according to Noyes et al. 1984). With $Ro = 6.0 \times 10^{-2}$, the filling factor is predicted to be $f_s = 0.7$. Thus the minimum values of $B_m$ in Table 6 lie below the equipartition field, as they do for the flare on Prox Cen (Poletto et al. 1988).

The parameters in Table 6 can be compared with those of large solar two-ribbon flares (See Bray et al. 1991; Pallavicini et al. 1990). Even the largest solar flares have areas which are less than about 0.2% of that of the solar hemisphere. With an area of less than 1% the main difference in the flare on CC Eri would be the high electron pressure implied (greater than $2 \times 10^{19}$ cm$^{-3}$ K, compared with values an order of magnitude lower in a solar flare). Alternatively, a pressure similar to that in a solar flare is found only with a substantially larger area (8% of the hemisphere) and larger height, than typical of a large solar flare. Thus in this model, which has continual energy input, the flare on CC Eri has either a higher density or a larger volume than a large solar two-ribbon flare. From this model alone any solution with an area between 1 and 10% of the stellar hemisphere gives physically possible parameters. In Section 5.1 we discuss an alternative model with *no* heating beyond the flare maximum, in attempt to restrict the possible solutions.

**Table 5.** Parameters of difference spectra (CEM RS-model).

| Spectral Group | $T_{max}$ ($10^6$ K) | $Em(V)_{max}$ ($10^{52}$ cm$^{-3}$) | $L_x^1$ | $\chi_r^2$ (24dof) |
|---|---|---|---|---|
| 3 | 14.8 ± 13.5 | 0.4 ± 0.6 | 5 ± 6 | 0.5 |
| 4 | 17.1 ± 4.8 | 1.1 ± 0.4 | 10 ± 3 | 1.6 |
| 5 | 21.9 ± 2.1 | 1.7 ± 0.4 | 14 ± 3 | 0.7 |
| 6 | 22.5 ± 2.1 | 3.1 ± 0.5 | 25 ± 4 | 1.0 |
| 7 | 27.5 ± 2.0 | 4.2 ± 0.5 | 30 ± 3 | 1.0 |
| 8 | 27.1 ± 1.4 | 4.1 ± 0.4 | 29 ± 2 | 1.4 |
| 9 | 22.0 ± 1.9 | 2.8 ± 0.4 | 23 ± 3 | 1.5 |
| 10 | 17.9 ± 4.7 | 2.1 ± 0.7 | 20 ± 6 | 1.1 |
| 11 | 13.8 ± 4.6 | 1.1 ± 0.8 | 13 ± 4 | 1.6 |
| 12 | 15.2 ± 3.5 | 1.1 ± 0.6 | 12 ± 7 | 1.0 |
| 13 | 17.2 ± 2.4 | 1.7 ± 0.4 | 16 ± 4 | 0.8 |
| 14 | 13.8 ± 4.0 | 0.9 ± 0.6 | 11 ± 7 | 0.6 |
| 15 | 11.6 ± 2.6 | 0.7 ± 0.2 | 8 ± 3 | 1.1 |
| 16 | 11.8 ± 4.3 | 0.6 ± 0.3 | 7 ± 4 | 1.1 |
| 17 | 6.9 ± 13.8 | 0.1 ± 0.3 | 1 ± 5 | 1.6 |
| 18 | 11.6 ± 13.1 | 0.1 ± 0.2 | 2 ± 3 | 0.6 |

[1] Energy release rate over the 0.2-2 keV band in unit of $10^{28}$ erg s$^{-1}$ after correction for interstellar absorption.

Table 6. Flare parameters of CC Eri.

| Region width (degree) | 33° | 20° | 10° | 5° |
|---|---|---|---|---|
| % of hemisphere affected | 8% | 3% | 1% | 0.2% |
| $B_m$ (Gauss) | 252 | 413 | 751 | 1474 |
| $H_m$ ($10^5$ km) | 2.42 | 1.46 | 0.71 | 0.38 |
| $V_{rise}$ (km s$^{-1}$) | 14.2 | 8.6 | 4.2 | 2.2 |
| $N_e^1$ ($10^{11}$ cm$^{-3}$) | 1.3 → 0.1 | 2.8 → 0.2 | 8.4 → 0.5 | 21.8 → 1.4 |

[1] The range in $N_e$ represents the values for the groups 7-18.

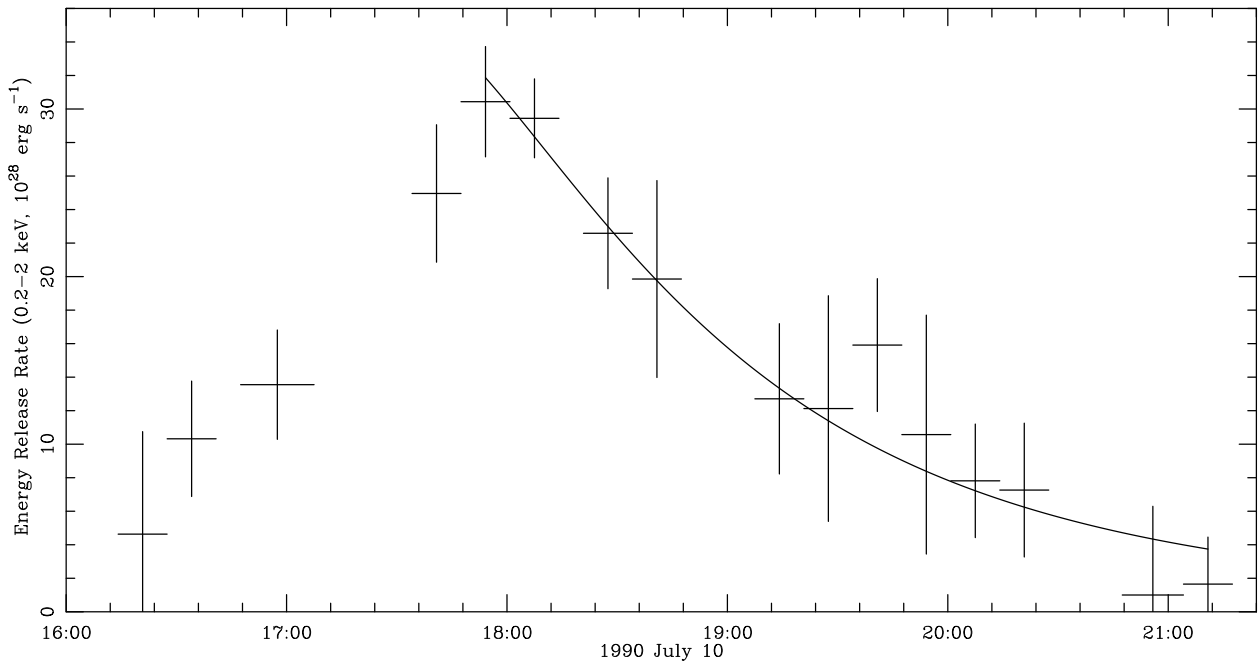

Figure 4. Evolution of the energy release rate of the two ribbon flare. The crosses represent the energy release rate measured by the PSPC and the solid line is the best fit curve of the magnetic reconnection model.

### 4.2 The 1990 July 10 Event – Modulation of Stellar Rotation?

As for other BY Dra variables, the optical photometric and spectroscopic peculiarities of CC Eri are attributed to spots on the star surface (Krzeminski 1969; Bopp & Evans 1973; Busko et al. 1977). The cause of the variation is the presence on the surface of magnetic starspots (in active regions) with a temperature a few hundred degrees cooler than the surrounding photosphere. It is expected that the starspots give rise to the optical light minimum and the associated active regions to the X-ray flux maximum. The starspot model, via a solar analogy, may provide a ready explanation for the X-ray activity observed with the PSPC. The X-ray flux increases and decreases when the active regions associated with the spots pass in and out of the line of sight. Based on this scenario we have created a starspot model to reproduce the X-ray light curve as observed with PSPC on 1990 July 10. Since the spot areas are relatively large, we assume that the active regions have the same size as the spots. In practice, this may underestimate the size of the active regions.

In a spherical coordinate system, $\gamma$, the angle between the line of sight and the normal line of an X-ray emitting area cell can be written as

$$\cos\gamma = \cos(\psi - \phi)\sin\theta\sin i + \cos\theta\cos i \qquad (4)$$

where $\phi$ is the stellar longitude and $\theta$ the co-latitude; $\psi$ is defined as the rotational phase and $i$ as the orbital inclination which is 42° (Evans 1959; Bopp & Evans 1973). The X-ray region is visible if the angle $\gamma$ is less than 90°.

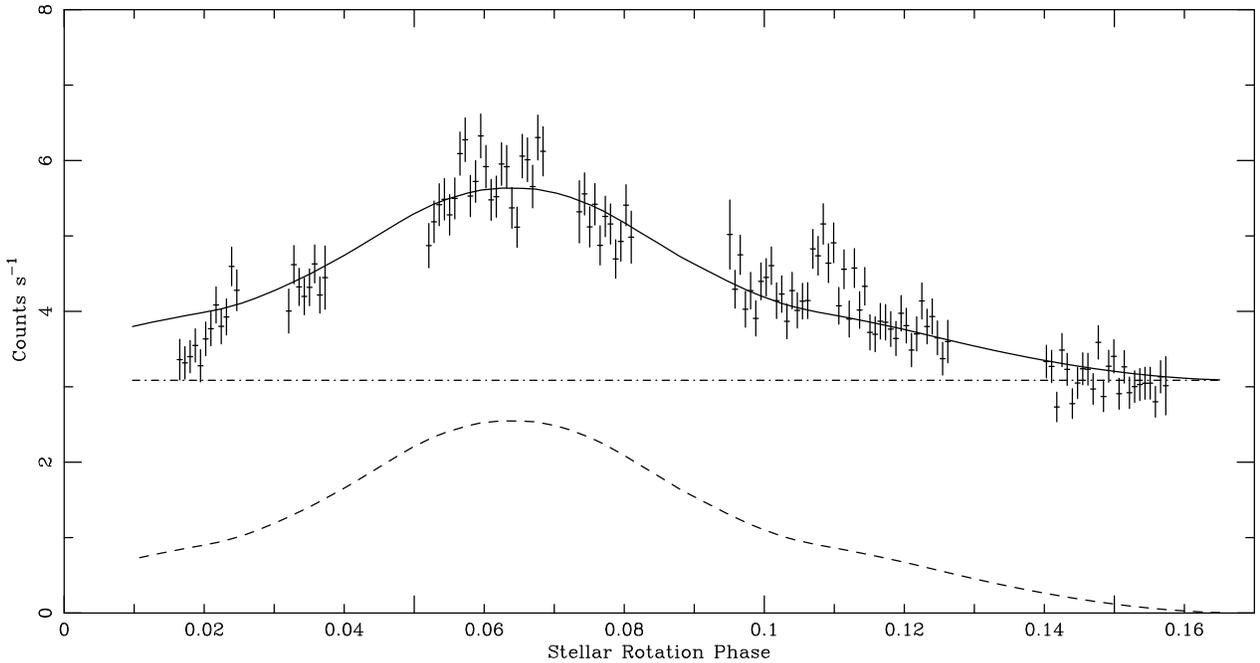

**Figure 5.** X-ray light curve of CC Eri as a function of the stellar rotation phase. The synthetic light curve from the starspot model is displayed as a solid line. The contributions of the north and south spots are shown by the dash-dot-dash line and by the long dash line respectively.

**Table 7.** Spot parameters of CC Eri.

|  | Longitude limits | Latitude limits | % of hemisphere affected | Spot X-ray flux erg cm$^{-2}$ s$^{-1}$ | Predicted Optical $L_{\min}/L_{\max}$ |
|---|---|---|---|---|---|
| Model 1 |  |  |  |  | 0.84 |
| North spot | 0° to 45° | −10° to 30° | 8 | $0.36 \times 10^9$ |  |
| South spot | 15° to 31° | −38° to −54° | 0.9 | $3.69 \times 10^9$ & $17.76 \times 10^9$ † |  |
| Model 2 |  |  |  |  | 0.89 |
| North spot | 6° to 36° | 10° to 40° | 4 | $0.77 \times 10^9$ |  |
| South spot | 15° to 31° | −38° to −54° | 0.9 | $3.69 \times 10^9$ & $17.76 \times 10^9$ † |  |
| Model 3 |  |  |  |  | 0.69 |
| North spot | 351° to 51° | 5° to 65° | 14 | $0.22 \times 10^9$ |  |
| South spot | 15° to 31° | −38° to −54° | 0.9 | $3.69 \times 10^9$ & $17.76 \times 10^9$ † |  |

† The surface flux of $17.76 \times 10^9$ erg cm$^{-2}$ s$^{-1}$ is in the region with longitude from 18° to 28° and latitude from −41.5° to −51.5°. Note that the fluxes required for the central part of the south spot exceed the stellar surface flux; see text for discussion.

Two starspots (which may be groups of spots), one located in the north hemisphere and another in the south, are required to reproduce the X-ray light curve on 1990 July 10. The north hemisphere spot has a uniform flux distribution, and the south spot has a flux distribution in which the flux is high in the central region ($\phi \sim 18° - 28°$, and $\theta \sim 131.5° - 141.5°$), low in the surrounding area. This region of concentrated flux is required to match the observed X-ray variation. In Table 7 we list the parameters of these spots. Each spot is divided into 100×100 X-ray emitting cells and the X-ray light curve is synthesized by adding the contribution of each cell according to equation (4). The shape of the X-ray light curve is largely determined by the location and size of the spot in latitude. The position of the X-ray flux maximum and minimum in the stellar rotation phase is a function of the longitude. We have adopted an ephemeris from Evans (1959) in which the stellar rotation period is 1.56145 days and the phase 0 is at JD2430000.0. The peak of the observed X-ray flux occurs at rotation phase $\psi = 0.06$. The longitude location of the spots is thus determined by maximizing equation (4). The size of the northern spot and its latitude location are not independently known and the three values used (see Table 7) are simply adopted from Bopp & Evans (1973) who modelled the optical light curves, over three periods of time, with these parameters. We have generated optical light curves using the same limb-darkening coefficients and ratio of spot temperature to photospheric temperatures as those used by Bopp & Evans (1973). The predicted ratios (see Table 7) of the optical minimum to maximum luminosities $L_{min}/L_{max}$ are consistent with the optical observations.

The X-ray light curve of CC Eri shown in Fig. 1 has been folded with the stellar rotation phase and is plotted in Fig. 5 together with the synthetic light curve and its two components. In the rotation phase 0-0.17, the size and location of the north spot has no effect on the shape of the X-ray light curve but gives a constant flux. The X-ray variation in this phase interval is completely determined by the location, size and the X-ray flux of the south spot. By contrast, the optical photometric variation is determined largely by the north spot. Although it is possible to choose parameters for the south spot that can reproduce the X-ray variations, as shown in Fig. 5, the solutions for the brightest region are *not* physically acceptable since even the radiation losses (0.2-2 keV band) alone exceed the stellar surface flux. Thus it seems *unlikely* that the variations observed are due to rotational modulation of activity related to starspot regions.

## 5 MODELLING FROM THE EMISSION MEASURES AND TEMPERATURES

Since emission measures and temperatures have been derived, some further modelling is possible. First, the change in the emission measure and temperature during the decay phase of the flare can be used to deduce geometrical parameters and limits to energy fluxes, under the assumption of *no* heating. This is an alternative model to that discussed in Section 4.1, which assumed continued energy input through magnetic reconnection. Secondly, the spectral groups before and after the flare can be used to investigate the *average* coronal conditions. The emission measure distribution derived from an energy balance model can be used to make comparisons with the value found at $1 - 2 \times 10^5$ K found from analyzing spectra obtained with the *IUE*, which are now in the public archive.

### 5.1 The Time-Varying Component

If the variation in the flux and temperature are interpreted as a flare, the difference spectra can be used to estimate either the length or the volume of the region. The full equation for the cooling rate, including a varying emission measure and temperature has been given by Veck et al. (1984). The conductive flux can be estimated by balancing it against the radiation below the region of maximum temperature. The equation for the change in $T_e$ then becomes

$$\frac{dT_e}{dt} = \frac{-7.2 \times 10^4 T_e^{1.25}/L - (0.49 Em(V)^{1/2}/V^{1/2} T_e)}{1 - (d \log(Em(V)/V)/3 \; d \log T_e)} \quad (5)$$

where $Em(V)$ is the volume emission measure for half the loop, $L$ is the loop half-length and $V$ is the half-loop volume. The denominator on the R.H.S. of equation (5) must be $> 0$. Using the data given in Table 5, $d \log Em(V)/3 \; d \log T_e$ satisfies this condition, between groups 7 to 11, and between groups 13 to 15. However, any increase in $V$ must be small, otherwise the denominator becomes $< 0$. Since the uncertainties in the individual values of $Em(V)$ and $T_e$ are large, we use only the changes in $Em(V)$ and $T_e$, and the average values, between (a) groups 8 to 11 and (b) 13 to 15, to examine the possible values of $L$ and $V$.

If $F_c(T_e) \gg F_R(T_e)$, then $L$ can be found, giving values of (a) $\sim 3.1 \times 10^{10}$ cm and (b) $\sim 1.8 \times 10^{10}$ cm. For comparison, the pressure-squared isothermal scale height at the mean temperatures are $\sim 4.0 \times 10^{10}$ cm and $\sim 2.8 \times 10^{10}$ cm. Only the product $N_e A^{1/2}$ ($A$ – area of one footpoint) can then be found and this has values of (a) $6.4 \times 10^{20}$ cm$^{-2}$ and (b) $5.5 \times 10^{20}$ cm$^{-2}$.

If $F_R(T_e) \gg F_c(T_e)$, then the volume can be found and the values are (a) $2V = 1.5 \times 10^{30}$ cm$^3$ and (b) $1.1 \times 10^{30}$ cm$^3$. (These can be compared with the volume of a spherically symmetric corona, of thickness corresponding to the pressure-squared isothermal scale height at the average quiescent temperature given in Section 5.2, i.e. $\sim 2 \times 10^{32}$ cm$^3$). In this case the density can also be found and the values are (a) $N_e \sim 1.3 \times 10^{11}$ cm$^{-3}$ and (b) $1.1 \times 10^{11}$ cm$^{-3}$, giving pressures comparable with those in a large solar flare.

Increasing the radiative term relative to the conductive term leads to larger values of $L$ and $V$, but lower values of $N_e A^{1/2}$ and $N_e$. Adding a heating term would, for the same values of $dT/dt$, lead to smaller values of $L$ and $V$.

Assuming that $F_c = F_R$ the total energy losses are (a) $5.3 \times 10^9$ erg cm$^{-2}$ s$^{-1}$ and (b) $2.6 \times 10^9$ erg cm$^{-2}$ s$^{-1}$, about a factor of three to five lower than the stellar surface flux ($1.3 \times 10^{10}$ erg cm$^{-2}$ s$^{-1}$).

Although the assumption of no heating may not be appropriate early in the decay, it is of interest to compare the parameters for groups 13 to 15, late in the flare, with those in the heated model, since in the latter case the heating decreases exponentially according to equation (3). Table 8 gives the parameters derived in the two models, for the case of $F_c > F_R$ since in the heated model it is assumed that only 10% of the total energy loss is in the form of radiation in the X-ray band. From Table 8 it can be seen that the two models agree best for the reconnection model which has an area of 8%. This area is within the range of star spot areas deduced by Bopp & Evans (1973), as given in Table 7.

### 5.2 Average Conditions

The mean coronal emission measure and temperature are derived from the ten spectral groups, 1 to 4 and 15 to 20. (Using a more restricted set 1 to 3 and 17 to 20 makes no significant difference). Studies of other stars with *ROSAT* and *EUVE* (Jordan et al. in preparation) suggest that of the various spectral fits, the higher temperature of a two temperature fit has the most physical significance. We adopt the results from the RS code. The mean values of the apparent emission measure and temperature are:

$$Em(V) = \int N_e^2 dV = 1.06 \times 10^{52} \text{ cm}^{-3} \quad (6)$$

and

$$T_e = 9.46 \times 10^6 \text{ K} \quad (7)$$

In practice, only some fraction of the emitting volume is observed, since the star shadows the rest. The fraction of a spherically symmetric corona that is observable is given by

$$G(r) = 0.5(1 + (1 - (R_\star/r)^2)^{1/2}) \quad (8)$$

**Table 8.** Comparison of parameters derived from the model with no heating and the reconnection model.

| | Model with no heating | | Reconnection model | | | |
| --- | --- | --- | --- | --- | --- | --- |
| | | | Area: 8% | | 3% | |
| | $L$ ($10^{10}$ cm) | $N_e A^{1/2}$ ($10^{20}$ cm$^{-2}$) | $L$ ($10^{10}$ cm) | $N_e A^{1/2}$ ($10^{20}$ cm$^{-2}$) | $L$ ($10^{10}$ cm) | $N_e A^{1/2}$ ($10^{20}$ cm$^{-2}$) |
| Groups 8-11 | 3.1 | 6.4 | 1.4 | 1.3 | 0.84 | 1.7 |
| Groups 13-15 | 1.8 | 5.5 | 2.0 | 7.4 | 1.2 | 9.5 |

Allowing also for the radial extent of the corona, the apparent emission measure as a function of $r$ is

$$Em(r)_{\rm APP} = \int N_e^2 (\frac{r}{R_\star})^2 G(r) dr = \frac{Em(V)}{4\pi R_\star^2} \quad (9)$$

The "true" emission measure is then

$$Em(r)_{\rm T} = \frac{Em(V)}{4\pi R_\star^2 G(r) f(r)} \quad (10)$$

where

$$f(r) = (\frac{r}{R_\star})^2 \quad (11)$$

In a plane parallel approximation to a uniform, non-extended corona, the value of $G(r)$ is 0.5, and $f(r)$ is 1.0, so that the *maximum* value of $Em(r)_{\rm T}$ is $9.67 \times 10^{29}$ cm$^{-5}$.

The electron pressure can then be found by writing $Em(r)_{\rm T}$ as

$$Em(r)_{\rm T} = P_e^2 \frac{H}{2} T_e^{-2} \quad (12)$$

where $P_e$ is in cm$^{-3}$ K and $H/2$ is the pressure-squared isothermal scale height, given by

$$\frac{H}{2} = 7.09 \times 10^7 T_e g_\star^{-1} \quad (13)$$

This simple approximation gives the *maximum* pressure and density in a *plane parallel* corona, since including the radial terms leads to lower values. (The pressures and densities would of course be even higher if the emitting area was restricted to only part of the corona, but there is not enough information to constrain such models).

In the plane parallel approximation the observed values of $Em(V)$ and $T_e$ lead to $P_e = 6.8 \times 10^{16}$ cm$^{-3}$ K and $N_e = 7.2 \times 10^9$ cm$^{-3}$. However, this approximation will not be entirely appropriate because the isothermal scale height is significant compared with the stellar radius. The average radius of the emitting corona can be estimated from $H/2$. (The height of the base of the corona is likely to be small, by analogy with the Sun). With $H/2 = 1.89 \times 10^{10}$ cm, one finds $(r/R_\star) \le 1.45$ and $G(r) \le 0.86$. Thus the *lower* limits to the true emission measure and the electron pressure are $Em(r)_{\rm T} \ge 2.7 \times 10^{29}$ cm$^{-5}$, and $P_e \ge 3.6 \times 10^{16}$ cm$^{-3}$ K.

The emission measure distribution below the coronal temperature is not known. It is possible to calculate the distribution by assuming that there is no non-thermal energy deposited below the corona, so that the radiation losses are balanced by the net conductive flux. In the spherically symmetric case the energy balance gives

$$\frac{1}{r^2} \frac{d(r^2 F_c(T_e))}{dr} = -0.87 \frac{P_e^2}{T_e^2} P_{\rm rad}(T_e) \quad (14)$$

where

$$F_c(T_e) = -\kappa_0 T_e^{5/2} \frac{dT_e}{dr} \quad (15)$$

and $\kappa_0 = 1.1 \times 10^{-6}$ erg cm$^{-1}$ K$^{-7/2}$ s$^{-1}$. The form of radiative power loss coefficient adopted is $P_{\rm rad}(T_e) = \varepsilon/T_e = 2.33 \times 10^{-16}/T_e$ erg cm$^3$ s$^{-1}$, to reproduce the RS code at $10^7$ K. Substituting

$$dT_e/dr = \frac{P_e^2}{T_e} \frac{1}{\sqrt{2} Em(0.3)} \quad (16)$$

where, $Em(0.3)$ is the average $Em(r)_{\rm T}$ over $\Delta \log T_e = 0.3$ dex, to match the typical region of line formation, leads to

$$d[(\frac{r}{R_\star})^2 \kappa_0 T_e^{3/2} \frac{P_e^2}{\sqrt{2} Em(0.3)}] = 0.87 (\frac{r}{R_\star})^2 \frac{\varepsilon \sqrt{2} Em(0.3) dT_e}{T_e^2} \quad (17)$$

Expressing $Em(0.3)$ in terms of the apparent emission measure, $Em(r)$, equation (17) can be written as

$$\frac{d \log(Em(r)/G(r))}{d \log T_e} = 4 \frac{d \log r}{d \log T_e} + \frac{3}{2} + 2 \frac{d \log P_e}{d \log T_e} - \frac{2\varepsilon}{\kappa_0} 0.87 (\frac{R_\star}{r})^4 \frac{Em^2(r)}{T_e^{5/2} G(r)} \frac{1}{P_e^2} \quad (18)$$

Our computer programme uses the apparent coronal emission measure, temperature and pressure to calculate $Em(r)$ in an iterative manner, making use of the subsidiary equation (16) and the equation of hydrostatic equilibrium, expressed as

$$\frac{dP_e}{dT_e} = -\frac{\mu m_{\rm p} g_\star \sqrt{2}}{k} \frac{Em(r)}{G(r)} \frac{1}{f(r)^2} \quad (19)$$

The radial factors are removed in the plane parallel version, and $G(r)$ is set to $1/2$. The absolute height scale is initially arbitrary.

The resulting emission measures are then used together with equations (16) and (19), to calculate the run of pressure and temperature with height above the chromosphere, using also, below $T_e = 2 \times 10^5$ K, emission measures derived from *IUE* spectra. Here we are concerned only with a comparison of the observed and calculated values of the emission measure at $T_e = 1 - 2 \times 10^5$ K. The full emission measure distribution and models of the chromosphere to the corona will be discussed in a later paper. The relevant emission measures are those found from the lines of C IV and N V. We have extracted 12 low resolution short wavelength IUE spectra from the archives to find the average surface fluxes. (See also Byrne et al. 1992). The emission measures were found as in Jordan et al. (1987), using the "solar" set of ion abundances from Jordan (1969), element abundances from Grevesse, Noels & Sauval (1992), the same collision strength for N V, but that for C IV reduced by a factor of 1.17. The resulting values of $Em(0.3)$ are $3.2 \times 10^{28}$ cm$^{-5}$ for C IV and $\le 6.6 \times 10^{28}$ cm$^{-5}$ for N V.

The plane parallel models are very sensitive to the coronal values of $Em(r)$ and $\varepsilon$ adopted when $Em(r)$ is near a critical maximum value. Adopting $Em(r) = 9.7 \times 10^{29}$ cm$^{-3}$, does *not* give an energy balance solution, but a slight reduction to $8.2 \times 10^{29}$ cm$^{-5}$ gives a solution with an emission measure of $3.6 \times 10^{28}$ cm$^{-5}$ at $2 \times 10^5$ K, similar to that derived from the *IUE* spectra. A small reduction in $\varepsilon$ would achieve the same result. The coronal pressure is $6.4 \times 10^{16}$ cm$^{-3}$ K and the coronal temperature (rounded to $10^7$) is reached by a height of $(h+R_\star) = 5.1 \times 10^{10}$ cm $(r/R_\star = 1.2)$.

In the spherically symmetric model the starting apparent emission measure is defined by equation (9), and the geometrical factors are then computed as part of the iterations. Because of the factor $f(r)$ (= 1.47) the true emission measure is smaller $(2.6 \times 10^{29}$ cm$^{-5})$, leading to a lower coronal electron pressure $(3.8 \times 10^{16}$ cm$^{-3}$ K), but the temperature gradients (from equation (16)) are similar. The model then has the same radial extent, but the emission measure produced at $T_e \sim 2 \times 10^5$ K is two orders of magnitude lower than that observed.

The plane parallel model is clearly not entirely self-consistent since the radial extent should be taken into account. But neither is spherically symmetric model satisfactory. This suggests that more complex geometries should be explored, but would require assumptions regarding area factors and the magnetic field geometry. However, the present models do limit the range of the coronal pressure to within about a factor of two. The values found are comparable with those in a well developed solar active region and are about two orders of magnitude lower than found in the CC Eri flare models. Observations with *EUVE* could in principal further constrain the emission measure distribution and thus the geometry.

The observed parameters can be compared with the predictions of Hearn's (1975) minimum energy loss (m.e.l.) hypothesis. The equations used are given in Montesinos & Jordan (1993) (equations 4.8 and 4.9). With $T_e = 9.46 \times 10^6$ K, the predicted emission measure and pressure are $Em(r) = 4.8 \times 10^{29}$ cm$^{-5}$ and $P_e = 4.7 \times 10^{16}$ cm$^{-3}$ K. These values lie between the 'observed' values with the plane parallel and spherically symmetric interpretations. As found by Montesinos & Jordan (1993) for a larger sample of main-sequence stars, the m.e.l. hypothesis gives a remarkably good prediction of average coronal properties.

The observed emission measure and temperature can also be used to estimate the energy lost from the corona by radiation and conduction. Using equation (4.18) and (4.21, or 4.22) from Montesinos & Jordan (1993), these are, respectively, $2.1 \times 10^7$ erg cm$^{-2}$ s$^{-1}$ and $1.0 \times 10^8$ erg cm$^{-2}$ s$^{-1}$, in the plane parallel interpretation. The total energy loss follows the trend with the Rossby number and $g_\star$, found by Montesinos & Jordan (1993) (equations 4.30). The radiation flux would be smaller in the spherically symmetric approximation. Thus even in the "average" corona, the energy loss are $\sim 1\%$ of the stellar surface flux, significantly larger than in the Sun.

Using the definition of the plasma $\beta$, the coronal magnetic field can be found from the emission measure and temperature (see equation (4.4) in Montesinos & Jordan 1993), giving $B_c = 21\beta^{-0.5}$ Gauss (plane parallel model).

## 6 CONCLUSIONS

CC Eri was observed in the periods 1990 July 9-11 and 1992 January 26-27 with the PSPC detector on board the *ROSAT* satellite. These high quality data give the first information on the temporal and spectral variability of CC Eri in the X-ray energy band.

During the *ROSAT* observations the X-ray intensity of the source is variable on timescales from a few minutes to hours. The X-ray luminosity is in the range $2.5 - 6.8 \times 10^{29}$ erg s$^{-1}$, which is similar to values found from previous measurements with the *Einstein* IPC and *EXOSAT* LE. On 1990 July 10 an X-ray flare-like event was detected with an exponential rise time of about one hour and a decay time of about two hours.

The X-ray spectrum of CC Eri can be well reproduced with either 2T or CEM (RS) models and the spectral results derived with either show the presence of high temperature ($\sim 10^7$ K) plasma around CC Eri. We find that the X-ray spectrum is also variable and the variations of the emission measure and to a lesser extent, the temperature, are correlated with the source intensity.

The X-ray variability of CC Eri observed with the PSPC may in principle be caused by flaring events and/or by the rotational modulation of active regions. However, the parameters required for the region causing X-ray variations are physically unrealistic. Thus it seems more likely that some sort of flare was observed. A model developed for solar two ribbon flares in which heating is produced by magnetic reconnection requires either that the flare has a larger volume and percentage area than a solar two ribbon flare, or higher densities. A comparison between the reconnection model parameters and an unheated model, towards the end of the flare, suggests that the solution with a larger area ($\sim 8\%$) and volume, and electron pressures similar to those in solar flares is the more likely. The larger flare area on CC Eri, compared with the Sun, may not be surprising given the larger optical star spot area. Thus a two ribbon flare provides one possible explanation of the variations observed. The occurrence of a flare would, by analogy with the Sun, require the presence of an active region, but the contribution of such a region to the 'quiescent' emission cannot be determined. If the non-flaring emission is attributed to an 'average' corona, then the electron pressure and emission measure are similar to those predicted by the minimum energy loss hypothesis, which gives an adequate description of a range of main-sequence stellar coronae. The electron pressure in the average corona is $\sim 4 - 6 \times 10^{16}$ cm$^{-3}$ K, comparable with that in a well-developed solar active region. The energy flux (radiation plus conduction) is $\sim 1\%$ of the stellar surface flux, substantially larger than for the Sun.

Ideally, CC Eri needs to be monitored simultaneously in optical, ultra-violet, and X-ray wavelengths over a complete binary cycle in future observations. Such observations could be used to study both flares and rotational modulation. Multi-wavelength observations could be used to map the X-ray distribution over the stellar disc and constrain the area covered by active regions. Observations with *EUVE* could be used to derive information on the emission measure distribution and hence place limits on the geometry of the emitting regions.

14  *H. C. Pan and C. Jordan*

**Acknowledgments**

We thank the UK *ROSAT* Data Archive Centre in University of Leicester for providing the PSPC data of CC Eri obtained during the *ROSAT* calibration and verification phase. We are also indebted to Dr. Andrea Prestwich in Smithsonian Astrophysical Observatory for assistance with the *Einstein* IPC data. We thank the referee Prof. R. Pallavicini for his helpful comments. HCP acknowledge the support of an SERC research assistantship under grant GR/H25539.